\documentclass[12pt]{iopart}
\usepackage{graphicx}
\usepackage{epsfig}

\begin{document}

\title{Control of quantum interference in the quantum eraser}
\author{L Neves, G Lima, J Aguirre, F A Torres-Ruiz, C Saavedra and A Delgado}
\address{Center for Quantum Optics and Quantum Information, Departamento de F\'{\i}sica,  Universidad de Concepci\'{o}n, Casilla 160-C, Concepci\'{o}n, Chile}
\ead{aldelgado@udec.cl}


\begin{abstract}
We have implemented an optical quantum eraser with the aim of studying this phenomenon in the context of state discrimination. An interfering single photon is entangled with another one serving as a which-path marker. As a consequence, the visibility of the interference as well as the which-path information are constrained by the overlap (measured by the inner product) between the which-path marker states, which in a more general situation are non-orthogonal. In order to perform which-path or quantum eraser measurements while analyzing non-orthogonal states, we resort to a probabilistic method for the unambiguous modification of the inner product between the two states of the which-path marker in a discrimination-like process. 
\end{abstract}

\pacs{03.65.Ud, 42.50.Xa, 42.50.Dv}

\submitto{\NJP}

\maketitle

\section{Introduction}

The principle of complementarity states that quantum systems have properties which are real but mutually exclusive \cite{Bohr28}. Basically it says that the wave-like and particle-like behaviours of a quantum system cannot be observed simultaneously in an experiment. A typical example is the Young's double slit experiment where it is impossible to obtain interference fringes and complete which-path (slit) information. Einstein tried to circumvent this statement by introducing the classic recoiling slit gedanken experiment: the position of the particle in a far screen and the momentum that it transfers to the double slit would be measured such that interference and which-path would be retrieved. However, using the uncertainty principle, Bohr refuted Einstein's idea, recovering the notion of complementarity \cite{Bohr83}. The role of entanglement in the wave-particle duality came later and was first discussed by Wootters and Zurek \cite{Wootters79} when analyzing Einstein's gedanken experiment. After this, Scully and Dr\"{u}hl reinforced this idea showing that, in some circumstances, the entanglement between interfering particles and the measurement device (which is a quantum system) is the reason for the loss of interference, and not the uncertainty principle \cite{Scully82}. The states of this device serve as which-path markers (WPM) of the possible paths of the interfering particle and if they are orthogonal, the interference is destroyed even if they are not measured. However, they have shown that this which-path information (WPI) can be ``erased'' and interference recovered by correlating measurements on the interfering particle with suitable measurements on the WPM. This phenomenon is referred to as {\it quantum eraser} \cite{Scully82,Scully89,Scully91}, and has been subject of intensive research in the last three decades (for a nice review see \cite{Aharonov05}).

Several quantum eraser experiments have been done employing two-photon states from spontaneous parametric down-conversion (SPDC) \cite{Herzog95,Tsegaye00,Kim00,Trifonov02,Walborn02,Gogo05,Scarcelli07}. A common feature of these implementations is the use of von~Neumann measurements to obtain WPI. Furthermore, with the exception of \cite{Trifonov02}, these experiments employ maximally entangled states among the interfering photon and the WPM.

In this article we report the implementation of an optical quantum eraser with the aim of studying this phenomenon in the context of state discrimination. Here a photon pair is produced by SPDC in an entangled polarization state and one photon of the pair is sent through a double slit. A single-photon interference pattern is observed at the far field, independent of the polarization state of the pair. In order to introduce WPI a quarter-wave plate is placed behind each slit with their fast axes orthogonally oriented [see Fig.~\ref{FIG0}(b)]. This birefringent double slit couples the polarization and spatial degrees of freedom of the interfering photon and thanks to the initial polarization entanglement the WPI becomes available into the polarization of both photons. Until this point there is a clear resemblance with the setup reported in Ref.~\cite{Walborn02}. However, the main distinguishing characteristics of our experiment are: First, we start with polarization entangled states with an arbitrary degree of entanglement, which can make the WPM to be non-orthogonal states. Second, we place a suitable polarization projector right after the birefringent double slit which makes the WPI to be carried just by the spatially separated photon. In \cite{Walborn02} the WPI is carried into two orthogonal maximally entangled polarization states and hence it can be erased by a suitable projection of the polarization of any photon of the pair, either the spatially separated photon or the interfering photon itself.
Finally, we use a polarization-sensitive Mach-Zehnder interferometer to manipulate locally and probabilistically the inner product of the WPM states, rather than resorting to polarization projections. With this setup the measurement of WPI or quantum erasure is closely related to the problem of discriminating among non-orthogonal states. The use of the interferometer will allow us to map, probabilistically, the initial states of the WPM onto pairs of states with any desired inner product at one of the output ports. Thus it is possible, for instance, to map initial non-orthogonal states onto orthogonal ones to get complete WPI, or onto collinear states to restore the maximum interference. Finally, is important to note that all the key ingredients for an optimal demonstration of a quantum eraser \cite{Kwiat94} are present: (i) possibility of delayed choice, i.e., the analysis of the WPM is carried out after the interfering particle has been detected \cite{Jacques08}, (ii) the setup employs single particles and (iii) the WPI is carried by a spatially separated system from the interfering particle.

This paper is organized as follows: in section~\ref{sec_theory} we describe the experimental setup and the theory behind our implementation of the quantum eraser. In section~\ref{sec_results} we present the experimental results and we conclude in section~\ref{sec_conclusion}.

\section{Experimental setup and theory}  \label{sec_theory}
\begin{figure}[t]
\hspace{2.5cm}
\rotatebox{-90}{\includegraphics[width=0.5\textwidth]{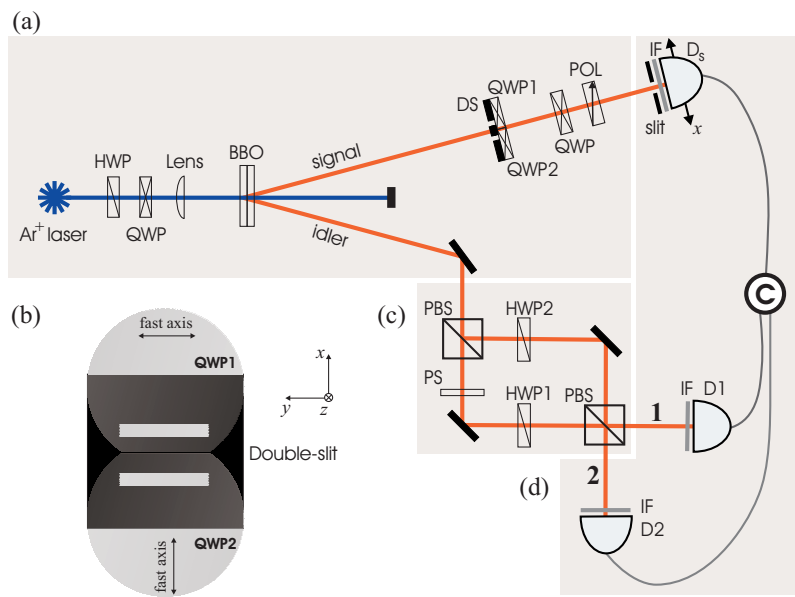} }
 \caption{Scheme of the experimental setup (see details in the text). HWP, half-wave plate; QWP, quarter-wave plate; DS, double slit; POL, linear polarizer; PBS, polarizing beam splitter; PS, phase shifter; IF, interference filter; D$_j$ ($j=\mathrm{s},1,2$), single photon detectors; C, single and coincidence counter. (a) state generation and state preparation stage;  (b) back-view of the double slit with QWPs, (c) polarization-sensitive Mach-Zehnder interferometer, and (d) detection stage.} 
 \label{FIG0} 
\end{figure}

The experimental setup is sketched in figure~\ref{FIG0}. A 351.1~nm single-mode Ar$^+$-ion laser pumps with 150~mW two 0.5~mm thick $\beta$-barium borate (BBO) nonlinear crystals cut for type-I phase matching and with their optical axes oriented $90^\circ$ with respect to each other. Pairs of photons, usually called signal (s) and idler (i), are generated by SPDC at an angle of $3^{\circ}$ with the pump beam and those with the same wavelength of 702.2~nm are selected by 10.0~nm bandwidth interference filters centered at this wavelength and placed in front of the photodetectors. The two-photon state generated  by this setup is given by \cite{Kwiat99} 
\begin{equation}
|\Psi\rangle=(a|H\rangle_\mathrm{s}|H\rangle_\mathrm{i}+b|V\rangle_\mathrm{s}|V\rangle_\mathrm{i})\otimes|\Psi_\mathrm{spa}\rangle,
\label{SOURCE}
\end{equation} 
where $H$ ($V$) denotes horizontal (vertical) polarization, $|a|^2+|b|^2=1$ and $|\Psi_\mathrm{spa}\rangle$ describes the spatial part of the two-photon state which will be discussed later. The amplitude and phase of $a$ and $b$ can be controlled by manipulating a half-wave plate (HWP) and a quarter-wave plate (QWP) placed in the pump beam [figure~\ref{FIG0}(a)]. For our future purposes the state in equation~(\ref{SOURCE}) can be cast in the following form:  
\begin{equation}
|\Psi\rangle=\frac{1}{\sqrt{2}}\left( |+\rangle_\mathrm{s}|\alpha_+\rangle_\mathrm{i}+|-\rangle_\mathrm{s}|\alpha_-\rangle_\mathrm{i}\right)\otimes|\Psi_\mathrm{spa}\rangle,
\label{SOURCE1}
\end{equation}
where 
\begin{eqnarray}
|\pm\rangle_\mathrm{s} & = & \frac{1}{\sqrt{2}}(|H\rangle_\mathrm{s}\pm|V\rangle_\mathrm{s}), \\
|\alpha_\pm\rangle_\mathrm{i} & = & a|H\rangle_\mathrm{i}\pm b|V\rangle_\mathrm{i},  \label{wpm1}
\end{eqnarray}
and the inner product of the later states is 
\begin{equation}
_\mathrm{i}\langle\alpha_+|\alpha_-\rangle_\mathrm{i} = |a|^2-|b|^2 \equiv \alpha,
\label{innerp}
\end{equation}
which turns out to be real.

\subsection{Observing single-photon interference}

After the photon pair has been generated, the signal photon is directed through a double slit  placed at 26~cm from the crystals. The spatial mode of the idler photon is defined by pinholes (1.5~mm diameter) placed along its path. In order to observe single photon interference, one of the basic requirements for a quantum eraser experiment \cite{Kwiat94}, the pump beam is focused into the BBO crystals using a 15~cm focal length lens [see figure~\ref{FIG0}(a)]. This procedure increases the transverse coherence length of the signal photon at the double slit plane \cite{Ribeiro94}.  The spatial part of the two-photon state right after the double slit can then be written as
\begin{equation}
|\Psi_\mathrm{spa}\rangle = \frac{1}{\sqrt{2}}(|\psi_1\rangle_\mathrm{s} + \rme^{\rmi\phi}|\psi_2\rangle_\mathrm{s})\otimes|\xi\rangle_\mathrm{i},
\label{spa}
\end{equation}
where $|\xi\rangle$ is the spatial mode of the idler photon defined by the pinholes; $|\psi_j\rangle$ is associated with the transmission of the signal photon through the slit $j=1,2$, and it is proportional to $\int dq\, \rme^{(-1)^j\rmi qd/2}\mathrm{sinc}(ql)|q\rangle$ \cite{Neves05} ($|q\rangle$ is a single-photon state with transverse component of the wave vector equal to $q$; $l$ and $d$ are defined below). The phase $\phi$ between the two possible paths can be set, for instance, by tilting the double slit. Using equations~(\ref{SOURCE1}) and (\ref{spa}) the total two-photon state is written (without normalization) as
\begin{equation}
|\Psi\rangle= \sum_{m=\pm}\left( |\psi_1,m\rangle_\mathrm{s}+\rme^{\rmi\phi}|\psi_2,m\rangle_\mathrm{s}\right)|\xi,\alpha_m\rangle_\mathrm{i}.
\label{SOURCE3}
\end{equation}

To show that (\ref{spa}) is in fact the spatial part of the two-photon state and that it is independent of the polarization degree of freedom we measure the coincidence and single count rates in the far field by scanning a ``pointlike'' detector at the signal arm and detecting the idler with a ``bucket'' detector, i.e., a detector that does not register where the photon has arrived. The measurements are done without analyzing their  polarization. This corresponds to tracing out the polarization of both photons (coincidences) or the idler photon and the signal polarization (single counts). In both cases the detection probability is \cite{Neves05}
\begin{equation}
I(x)\propto \mathrm{sinc}^2(klx/z)[1+\cos(kdx/z+\phi)],
\label{interf1}
\end{equation}
which is a typical double slit interference pattern. $k$ is the pump wave number, $l$ is the slit half width (40~$\mu$m), $d$ is the slits centre-to-centre separation (280~$\mu$m), $x$ is the transverse position of the signal detector and $z$ is the distance from the double slit to the detector (in this experiment $z$ is actually the focal length of a lens). The basic setup is shown in figure~\ref{FIG0}(a), but in this case there is neither quarter-wave plates (QWP) behind the double slit nor polarization projector (QWP+Polarizer) after that. Also, the idler is sent directly to a detector without passing through the interferometer. Both detectors are placed in the focal plane of a 30~cm focal length lens.  In front of the signal detector there is a 50~$\mu$m $\times$ 5~mm slit oriented parallel to the double slit while at the idler detector a 3~mm pinhole. 

\begin{figure}[t]
 \centering 
\includegraphics[width=0.6\textwidth]{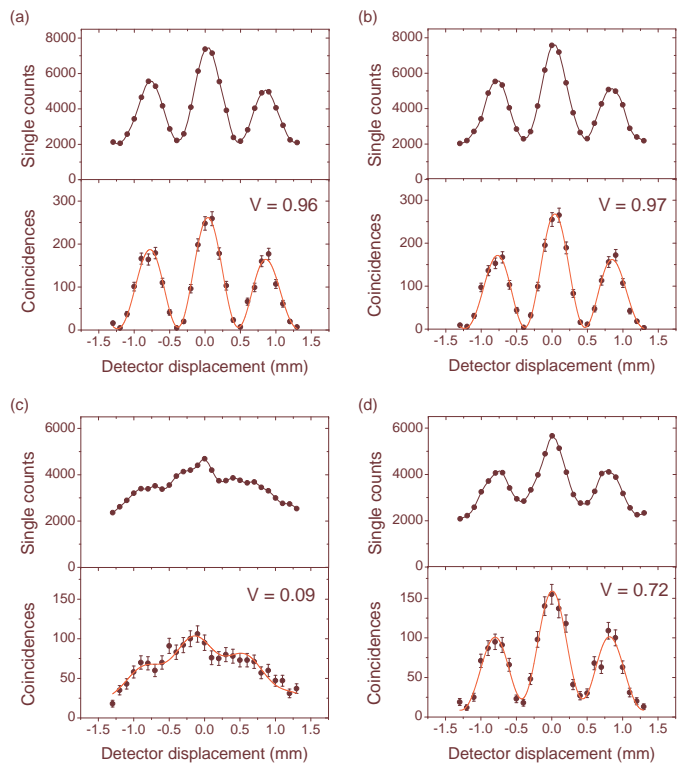} 
\vspace{-2.5cm}
 \caption{Signal count rates and coincidences ($\bullet$) in the far field for a double slit \emph{without} quarter-wave plates [(a) and (b)] and \emph{with} quarter-wave plates [(c) and (d)]. In (a) and (c) the entanglement in polarization is maximal while it is non-maximal in (b) and (d). The solid lines for the singles are just a guide while for the coincidences they are theoretical fits obtained from equation~(\ref{INTERFERENCESIGNAL}) using a normalization and the visibility (shown in the insets) as a free parameters. The visibilities of the singles are the same when the background noise is subtracted. Different count rates resulted from the losses due to the wave plates and the non-identical alignment in the two cases.} 
 \label{singles} 
\end{figure}

We prepare a maximally and a non-maximally entangled polarization state and characterize them by quantum state tomography \cite{James01}. The purities obtained were higher than 95\%, which allowed us to consider the two-photon polarization state to be essentially pure and given by equation~(\ref{SOURCE}). In these measurements we set $\phi$ in (\ref{interf1}) to be zero. For the maximally entangled state the inner product of $\{|\alpha_\pm\rangle_\mathrm{i}\}$ [see equation~(\ref{innerp})] is $\alpha = 0.02\pm 0.01$ and the results for the singles and coincidences are shown in figure~\ref{singles}(a). For the non-maximally entangled we get $\alpha = 0.70\pm 0.03$ and the results are shown in figure~\ref{singles}(b). The less-than-one visibilities are due to the finite size of the signal detector and the finite size of the source which makes the degree of transverse coherence differs from the unity. Nevertheless, it is clear from these graphics that the spatial part of the two-photon state is well described by equation~(\ref{spa}) and that there is no coupling with the polarization degree of freedom.

\subsection{Introducing which-path information}

At this point there is no WPI available yet. In order to introduce it we place a quarter-wave plate behind each slit with their fast axes orthogonally oriented as shown in figure~\ref{FIG0}(b). The action of these wave plates couples the polarization and spatial degrees of freedom of the signal photon and transforms the two-photon state given by equation~(\ref{SOURCE3}) into
\begin{eqnarray}
|\Psi\rangle & = & \left(|\psi_1\rangle_\mathrm{s}|\alpha_+\rangle_\mathrm{i}+
\rme^{\rmi\phi}|\psi_2\rangle_\mathrm{s}|\alpha_-\rangle_\mathrm{i}\right)|\xi\rangle_\mathrm{i}|L\rangle_\mathrm{s} \nonumber\\
 & & \mbox{} + \left(|\psi_1\rangle_\mathrm{s}|\alpha_-\rangle_\mathrm{i}+\rme^{\rmi\phi}|\psi_2\rangle_\mathrm{s}
|\alpha_+\rangle_\mathrm{i}\right)|\xi\rangle_\mathrm{i}|R\rangle_\mathrm{s},
\label{psi1}
\end{eqnarray}
where $L$ and $R$ represent left and right circular polarizations, respectively. This state already carries WPI in the \emph{two-photon} polarization state due to the initial polarization entanglement, what has been shown in \cite{Walborn02}. 

Using the same states we made the same set of measurements described previously. For the initial maximally entangled polarization state the results are shown in figure~\ref{singles}(c). As the WPM states are (ideally) orthogonal the interference observed in figure~\ref{singles}(a) is destroyed. The residual visibility observed in this case ($0.09\pm 0.03$) is due to the non-perfect alignment of the wave plates behind the slits which produces states $\{|\alpha_\pm\rangle\}$ non-exactly orthogonal. For the initial non-maximally entangled state, the WPM states are non-orthogonal and the visibility of the pattern observed in figure~\ref{singles}(b) is reduced to a degree that depends on their inner product. In this case $\alpha=0.70\pm 0.03$ and as shown in figure~\ref{singles}(d) the visibility is $0.72\pm 0.04$ which agrees well with the inner product within the experimental error. 

Some of the basic requirements for an optimal demonstration of a quantum eraser have been fulfilled, that is, we have single photon interference and we can imprint WPI which can be either used to detect which-path or erased to restore interference. However, as stressed in \cite{Kwiat94} it is pedagogically preferable that the WPI be carried separately from the interfering particle. This is not the case for the state in equation~(\ref{psi1}). To satisfy this condition we must make a further projection of the signal polarization right after its transmission through the birefringent double slit. Projecting, for example, onto $L$ the two-photon state in (\ref{psi1}) becomes
\begin{equation}
|\Psi\rangle=\frac{1}{\sqrt{N}}\left(|\psi_1\rangle_\mathrm{s}|\alpha_+\rangle_\mathrm{i}+
\rme^{\rmi\phi}|\psi_2\rangle_\mathrm{s}|\alpha_-\rangle_\mathrm{i}\right),
\label{SOURCE3-FILTERED}
\end{equation}
where $N=2\left[1+\mathrm{Re}(\rme^{\rmi\phi}\alpha\langle\psi_1|\psi_2\rangle)\right]$ is a normalization constant. In this expression we have omitted the factorable polarization (spatial) state of the signal (idler) photon to stress that the interfering particle (signal) and the WPM (idler) are nonlocally correlated. This projection just forces the entanglement between signal spatial degree of freedom and idler polarization and in this sense works as a \emph{disentanglement eraser} \cite{Garisto99}. Besides the pedagogical appeal, working with this state will allow us to manipulate locally the state of the WPM and study the quantum eraser in the context of state discrimination. 
 
 After tracing out the polarization or the idler in (\ref{SOURCE3-FILTERED}), the coincidence or single counts detection probability will be given by 
\begin{equation}
I(x)\propto \mathrm{sinc}^2(klx/z)[1+\alpha\cos(kdx/z+\phi)].
\label{INTERFERENCESIGNAL}
\end{equation}
Comparing with equation~(\ref{interf1}) one can see that the visibility that was one in that case is now governed by the absolute value of the inner product, $\alpha$, of the WPM states given by (\ref{innerp}) and shown in the results of figure~\ref{singles}. The phase of this pattern depends on the sign of $\alpha$.

\subsection{Which-path and quantum eraser measurements}

In order to make which-path or quantum eraser measurements the WPM must be projected onto a suitable state and its detection be correlated to the detection of the interfering particle, which in our case is done through coincidence measurements. From equation~(\ref{SOURCE3-FILTERED}) one can see that when the entanglement among these systems is maximal, WPI can be retrieved by measuring an observable spanned by the \emph{orthogonal} states $\{|\alpha_\pm\rangle_\mathrm{i}\}$. In this case the signal count rate conditioned upon the detection of the idler do not exhibit interference. Actually, as a consequence of the entanglement the interference is destroyed even when the WPM is not detected or it is detected by a polarization insensitive measurement as is shown in figures~\ref{singles}(a) and (c). On the other hand, WPI can be erased and interference restored by measuring an observable whose eigenstates has equal components on the states $\{|\alpha_\pm\rangle_\mathrm{i}\}$ and then correlating the results to the detection of the interfering particle. For instance, when the idler is projected onto the symmetric $(|\alpha_+\rangle_\mathrm{i}+|\alpha_-\rangle_\mathrm{i})/\sqrt{2}$ or antisymmetric $(|\alpha_+\rangle_\mathrm{i}-|\alpha_-\rangle_\mathrm{i})/\sqrt{2}$ state and its detection is correlated to the detection of the signal photon, the interference is restored in the form of fringes or antifringes, respectively.

A more general situation corresponds to non-maximal entanglement in (\ref{SOURCE3-FILTERED}), or equivalently to a WPM described by two non-orthogonal states. In this case the WPI will be partial which reduces the visibility of the interference pattern as can be seen in equation~(\ref{INTERFERENCESIGNAL}) and in figures~\ref{singles}(b) and (d). This partial WPI can be erased in the same way as in the case of a maximally entangled state. However, the intensity of each pattern will be modulated by $|a|^2$ (fringes) and $|b|^2$ (antifringes). The retrieval of WPI in this case is a more delicate matter. Non-orthogonal states cannot be perfect and deterministically discriminated  \cite{Peres88} since the eigenstates of any observable acting on the Hilbert space has non-vanishing components on both states $\{|\alpha_\pm\rangle_\mathrm{i}\}$. The best strategy  of error-free identification turns out to be of probabilistic nature and considers the possibility of an inconclusive identification \cite{Cheffles98,Cheffles982}.

\subsection{Controlling the which-path marker}

For arbitrary states of the WPM we can resort to a \emph{probabilistic} method for the unambiguous modification of their inner product. In our experiment this is done by using a polarization-sensitive Mach-Zehnder interferometer depicted in figure~\ref{FIG0}(c). The first polarizing beam splitter (PBS) splits the input beam into two propagation paths ($H$ polarization is transmitted and $V$ is reflected). In each path a half-wave plate rotates the polarization depending on its angle ($\gamma_1$ and $\gamma_2$) and the paths are then recombined at a second PBS. To understand the role of the interferometer onto the WPM states let us suppose that a single photon in the state $|\alpha_{\pm}\rangle$ [equation~(\ref{wpm1})] enters through the upper port as indicated in figure~\ref{FIG0}(c). Assuming that the pathlength difference between the two arms
is zero, the following unitary transformation, $U_I$, is performed: 
\begin{equation}
U_I |\alpha_\pm\rangle=N_1|\alpha^{(1)}_\pm\rangle-\rmi N_2 |\alpha^{(2)}_\pm\rangle,
\label{INTERFEROMETER}
\end{equation}
where
\begin{eqnarray}   \label{alphaWPM1}
|\alpha_{\pm}^{(1)}\rangle \! &=& \! \frac{1}{\sqrt{N_1}}\left[a\cos(2\gamma_1)|H\rangle\pm b\cos(2\gamma_2)|V\rangle\right]|1\rangle,
\\
|\alpha_{\pm}^{(2)}\rangle \! &=& \! \frac{1}{\sqrt{N_2}}\left[a\sin(2\gamma_1)|V\rangle \pm b\sin(2\gamma_2)|H\rangle\right]|2\rangle,
\label{ALPHAO}
\end{eqnarray}
$|1\rangle$ and $|2\rangle$ indicate the output ports $1$ and $2$, respectively, and $N_1$ and $N_2$ are the probabilities that an input photon will exit through the respective port. They are given by
\begin{eqnarray} \label{prob1}
N_1&=&|a|^2\cos^2(2\gamma_1)+|b|^2\cos^2(2\gamma_2),
\\
N_2&=&|a|^2\sin^2(2\gamma_1)+|b|^2\sin^2(2\gamma_2).
\label{prob2}
\end{eqnarray}

The two-photon state in equation~(\ref{SOURCE3-FILTERED}) when the idler goes through this interferometer is obtained by applying the transformation $U_I$ onto $|\Psi\rangle$, that gives
\begin{equation}
U_I|\Psi\rangle=\sqrt{\frac{N_1M_1}{N}}|\Phi_1\rangle-\rmi \sqrt{\frac{N_2M_2}{N}}|\Phi_2\rangle,
\label{AFTERINTERFEROMETER}
\end{equation}
where the orthogonal two-photon states $|\Phi_j\rangle$ (for $j=1,2$)  are 
\begin{eqnarray}  \label{PHI1}
|\Phi_j\rangle=\frac{1}{\sqrt{M_j}}\left( |\psi_1\rangle_\mathrm{s}|\alpha_+^{(j)}\rangle_\mathrm{i}
+\rme^{\rmi\phi}
|\psi_2\rangle_\mathrm{s}|\alpha_-^{(j)}\rangle_\mathrm{i}\right),
\end{eqnarray}
with $M_j=2\left[1+\mathrm{Re}(\rme^{\rmi\phi}\langle\psi_1|\psi_2\rangle\langle\alpha_+^{(j)}|\alpha_-^{(j)}\rangle)\right]$.
The transformation implemented by the interferometer maps the initial states of the WPM whose inner product is given by equation~(\ref{wpm1}) onto states in each output port with the inner products, $\langle\alpha_+^{(j)}|\alpha_-^{(j)}\rangle$, given by
\begin{equation}
\alpha^{(1)}(\Psi_{\mathrm{pol}},\gamma_1,\gamma_2)=
\frac{|a|^2\cos^2(2\gamma_1)-|b|^2\cos^2(2\gamma_2)}
{|a|^2\cos^2(2\gamma_1)+|b|^2\cos^2(2\gamma_2)},
\label{ALPHA1}
\end{equation}
and
\begin{equation}
\alpha^{(2)}(\Psi_{\mathrm{pol}},\gamma_1,\gamma_2)=\alpha^{(1)}(\Psi_{\mathrm{pol}},\pi/4-\gamma_1,\pi/4-\gamma_2)
\label{ALPHA2},
\end{equation}
with probabilities $(\ref{prob1})$ and $(\ref{prob2})$, respectively. ($\Psi_{\mathrm{pol}}$ denotes the initial polarization two-photon state, through the coefficients $a$ and $b$.) For instance, the angles $\gamma_1$ and $\gamma_2$ can be set to transform the initial non-orthogonal states of the WPM into orthogonal states at a given output port with a certain probability. In this particular case the interferometer implements the optimal unambiguous discrimination of the states $\{|\alpha_{\pm}\rangle_\mathrm{i}\}$ \cite{Torres08}. 

Since the action of the interferometer is unitary and local, that is, it affects only the idler photon, the inner product of the WPM states is preserved, i.e.,
\begin{equation}
\alpha = N_1\alpha^{(1)}(\Psi_{\mathrm{pol}},\gamma_1,\gamma_2)+N_2\alpha^{(2)}(\Psi_{\mathrm{pol}},\gamma_1,\gamma_2).
\label{cnorm}
\end{equation}
In the same way the signal detection probability in equation~(\ref{INTERFERENCESIGNAL}) does not change. It is determined only by the initial polarization entanglement. However, the signal detection probability conditioned upon the detection of the idler (without analyzing its polarization) at output ports $1$ or $2$ produces interference patterns, $I^{(1)}(x)$ or $I^{(2)}(x)$, respectively, which depend on the action of the interferometer. They are given by 
\begin{equation}
I^{(j)}(x)  \propto  \mathrm{sinc}^2(klx/z)[1+\alpha^{(j)}(\Psi_{\mathrm{pol}},\gamma_1,\gamma_2)\cos(kdx/z+\phi)],
\label{PATTERNS}
\end{equation}
for $j=1,2$. The weighted sum of these two conditional interference patterns gives a pattern equal to that of equation~(\ref{INTERFERENCESIGNAL}), which can be seen by using  equation~(\ref{cnorm}).

\section{Experimental results}   \label{sec_results}

To perform the experiment described in the previous section and sketched in figure~\ref{FIG0} we place a quarter-wave plate followed by a linear polarizer right after the birefringent double slit to project the signal polarization onto $L$ and produce the state given by equation~(\ref{SOURCE3-FILTERED}) \cite{Comment2}. The signal pointlike detector is placed at 20~cm from the double slit. (In this experiment the lens in the signal and idler arm has not been used.) The idler spatial mode is selected by 1.0~mm pinholes and it goes through the interferometer which implement the probabilistic modification of the inner product of the WPM states.  The optical path difference between the two arms, within the coherence length of the idler photon, is adjusted by rotating a 1-mm-thick glass plate inserted in the path 1 [PS in figure~\ref{FIG0}(c)]. A single photon detector is placed at the output port 1 and in front of it there is a 3~mm pinhole. All the coincidence measurements between $D_\mathrm{s}$ and $D_1$ are performed with a specific configuration of the half-wave plates in the interferometer and with the signal detector scanning in the $x$-direction.  The measurements at output port 2 have not been done simultaneously with the measurements at 1. We just direct the light from 2 to the same detector placed at output port 1 instead. Single and coincidence counts are registered in a counter (C) with a resolving time of 5~ns.

\begin{figure}[t]
 \centering 
 \includegraphics[width=0.65\textwidth]{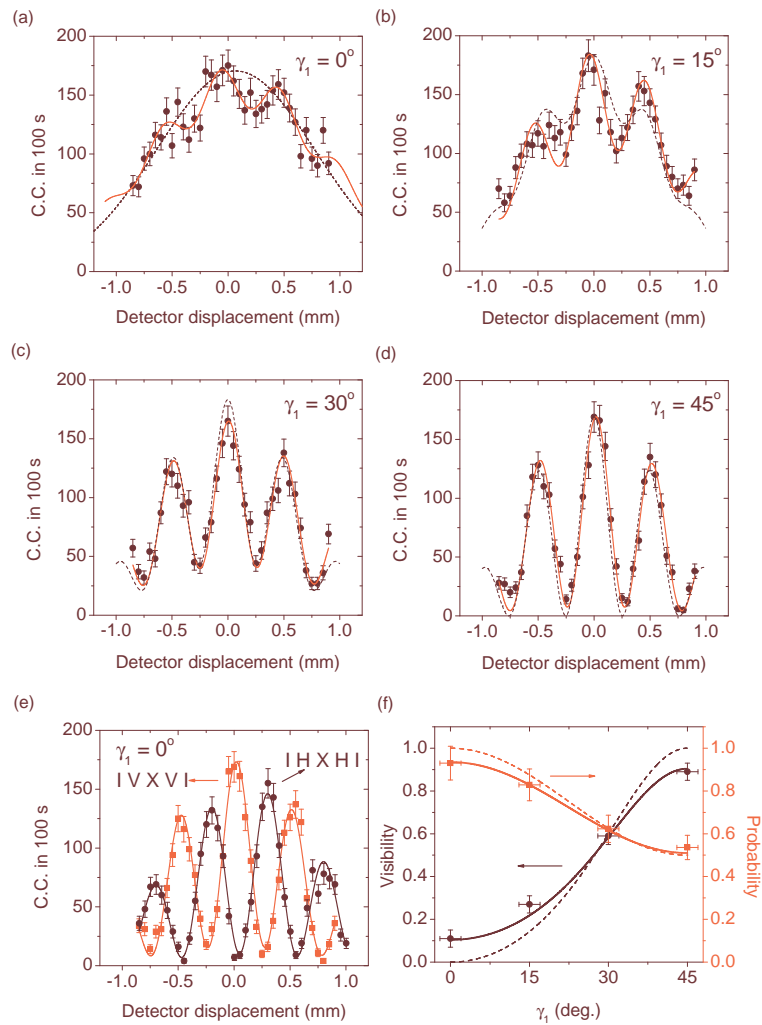} 
 \caption{Experimental results ($\bullet$) for an initial maximally entangled polarization state. Coincidence counts (C.C.) were recorded by scanning the signal detector in the $x$-direction and detecting idler photons at output port 1 (figure~\ref{FIG0}) keeping  $\gamma_2=0^\circ$ and (a) $\gamma_1=0^\circ$, (b) $\gamma_1=15^\circ$, (c)  $\gamma_1=30^\circ$and (d) $\gamma_1=45^\circ$. (e) Erasing the WPI and restoring the interference for the experiment in (a). The fringe (antifringe) is obtained by projecting the idler photon onto $|V\rangle\langle V|$ ($|H\rangle\langle H|$). (f) Visibility and probability versus HWP1 angle for the measurements (a)--(d). Dashed lines correspond to the theoretical prediction based on equation~(\ref{PATTERNS}) for interference patterns, (\ref{prob1}) for probability and the absolute value of (\ref{ALPHA1}) for the visibility. Solid lines are the theoretical predictions with corrections described in the text.}
 \label{FIG2} 
\end{figure}

\begin{figure}[t]
 \centering 
 \includegraphics[width=0.65\textwidth]{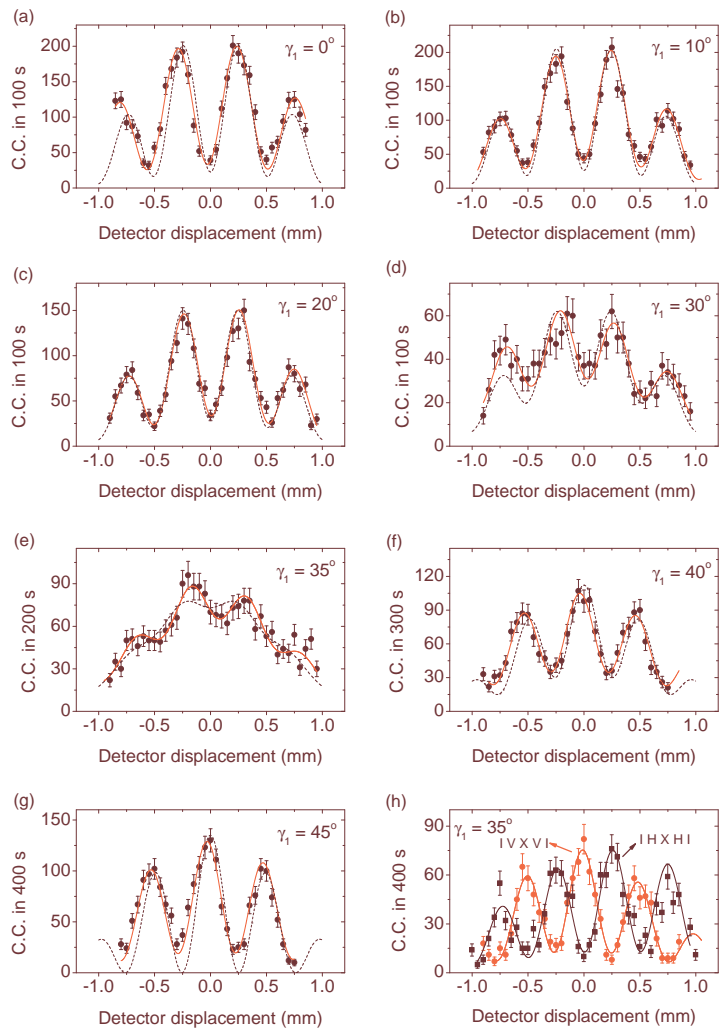} 
  \caption{Experimental results ($\bullet$) for an initial non-maximally entangled polarization state. Coincidence counts (C.C.) were recorded by scanning the signal detector in the $x$-direction and detecting idler photons at output port 1 (figure~\ref{FIG0}) keeping  $\gamma_2=20^\circ$ and (a) $\gamma_1=0^\circ$, (b) $\gamma_1=10^\circ$, (c)  $\gamma_1=20^\circ$, (d) $\gamma_1=30^\circ$, (e) $\gamma_1=35^\circ$, (f) $\gamma_1=40^\circ$ and (g) $\gamma_1=45^\circ$. (h) Erasing the WPI and restoring the interference for the experiment in (e). The fringe (antifringe) is obtained by projecting the idler photon onto $|V\rangle\langle V|$ ($|H\rangle\langle H|$). Dashed lines correspond to the theoretical prediction based on equation~(\ref{PATTERNS}). Solid lines are the theoretical predictions with corrections described in the text.} 
\label{FIG1} 
\end{figure}

\begin{figure}[t]
\hspace{2.5cm}
\rotatebox{-90}{\includegraphics[width=0.6\textwidth]{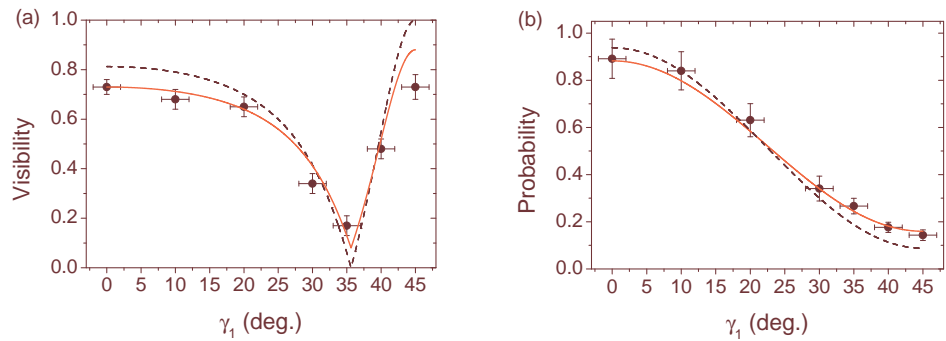} }
\vskip -4.5cm
 \caption{(a) Visibility and (b) probability versus HWP1 angle for the measurements (a)--(g) in figure~\ref{FIG1}. Dashed lines correspond to the theoretical prediction based on equation~(\ref{prob1}) for probability and the absolute value of (\ref{ALPHA1}) for the visibility. Solid lines are the theoretical predictions with corrections described in the text.} 
 \label{FIG4} 
\end{figure}

In the following we show the experimental data and analyze them with the help of the theoretical expressions of section~\ref{sec_theory}. The results are shown in figures~\ref{FIG2}--\ref{exit2}. In all graphics the dashed lines correspond to the theoretical predictions considering the setting of the angles $\gamma_1$ and $\gamma_2$ of the half-wave plates 1 and 2, respectively, together with the measured values of $a$ and $b$. Here we do not assume any correction. On the other hand, the solid lines for the visibility and probability curves correspond to the theoretical fits using the same parameters and taking into account that the maximal experimentally obtained visibility with the birefringent double slit was $0.9$ (for a polarization product state) while the minimum was $0.09$ [see figure~\ref{singles}(c)]. The reasons for that are the finite size of the signal detector, the less-than-one degree of transverse coherence of the source and mainly the non-perfect alignment of the quarter-wave plates 1 and 2. We also assume a small error in the $L$ polarization projection which leaves a small percentage of the component with $R$ polarization in the two-photon state [see equation~(\ref{psi1})]. For the interference patterns, the fits (solid lines) were obtained by using equation~(\ref{PATTERNS}) and taking as free parameters a normalization constant, the visibility and a phase offset. It is clear from all figures that the fit with corrections provides a better agreement between theory and experimental results.

\subsection{Orthogonal which-path marker states}

Figure~\ref{FIG2} shows the results for an initial maximally entangled polarization state. In this experiment the coincidences between signal and idler were measured with idler photons exiting through output port 1 only. The angle of the half-wave plate 2 ($\gamma_2$) was kept fixed at $0^{\circ}$ while the angle of the half-wave plate 1 ($\gamma_1$) changed from one measurement to another. For $\gamma_1=0^{\circ}$ the interferometer does not act on the states of the WPM. Since in this case these states are nearly orthogonal the interference is nearly destroyed even when the idler polarization is traced out. This is shown in figure~\ref{FIG2}(a) where the visibility of the pattern is $0.11\pm 0.04$. Any change of $\gamma_1$ will only increase the absolute value of $\alpha^{(1)}(\Psi_{\mathrm{pol}},\gamma_1,\gamma_2)$ [see equation~(\ref{ALPHA1})]. Consequently, the WPI decreases and the interference starts to show up: the intermediate cases are shown in figures~\ref{FIG2}(b) and (c). The complete erasure of WPI and restoration of the interference with maximum visibility is reached when $\gamma_1=45^{\circ}$, in which case the state of the WPM is $|V\rangle$. This is shown in figure~\ref{FIG2}(d) with a visibility of $0.89\pm 0.04$ which is close to the maximum value achievable in this setup.  The visibilities of the interference patterns as well as the probabilities that the idler exit through port 1 for the experiments of figures~\ref{FIG2}(a)--(d) is shown in figures~\ref{FIG2}(f) as function of $\gamma_1$. 

For the experiment of figure~\ref{FIG2}(a) the WPI can also be erased and interference restored via polarization projection \cite{Walborn02}. When the arm 1 (2) of the interferometer is blocked, idler polarization is projected onto $V$ ($H$) which leads to an interference pattern in the form of fringes (antifringes), as $\phi=\pi$ in equation~(\ref{SOURCE3-FILTERED}) \cite{Comment2}. The results are shown in figure~\ref{FIG2}(e). Note that the averaged sum of these two  patterns gives an interference pattern roughly equal to that of figure~\ref{FIG2}(a).

This particular experiment is similar to that of \cite{Walborn02} although in our case it is not necessary to include or remove a polarizer in the idler path to access all the  situations shown in figures~\ref{FIG2}(a)--(d) and we also have access to the idler exiting through output port 2, which is not possible by using just a polarizer. The main differences arise when the initial states of the WPM are non-orthogonal, which will be discussed next.

\subsection{Non-orthogonal which-path marker states}

For this experiment we used the same initial non-maximally entangled polarization state as in section~\ref{sec_theory}. The coefficients of this state are $|a|=0.92$ and $|b|=0.38$ which gives $\alpha=0.7$ [see figure~\ref{singles}(d)]. The interference patterns (in coincidence) were obtained with the angle $\gamma_2$ kept fixed at $20^{\circ}$ and different values of $\gamma_1$.  Figure~\ref{FIG1} shows these patterns for measurements at output port 1 with the angles of the half-wave plate 1 shown in the insets. The graphics (a) to (e) show interference antifringes because the inner products [equation~(\ref{ALPHA1})] are non-negative for the values of the parameters $a$, $b$, $\gamma_1$ and $\gamma_2$ while $\phi=\pi$ in equation~(\ref{PATTERNS}) \cite{Comment2}. On the other hand, the patterns (f) and (g) exhibit fringes as a consequence of the negative values of (\ref{ALPHA1}) for those values of $\gamma_1$.

The theoretical behaviour of the visibility (for the used parameters) is shown in the dashed line of figure~\ref{FIG4}(a). For $\gamma_1\leq 20^{\circ}$ it is bigger than or equal to the initial value of 0.7 (equality holds for $\gamma_1=\gamma_2$). As $\gamma_1$ increases the visibility decreases until reach zero at $\gamma_1=35.6^{\circ}$ and starts to increase again until 0.7 at $\gamma_1=41^{\circ}$. Thus in the range $20^{\circ}<\gamma_1< 41^{\circ}$ the interferometer maps the initial non-orthogonal states of the WPM onto states $\{|\alpha_{\pm}^{(1)}\rangle_\mathrm{i}\}$ with smaller absolute values of the inner product. In particular, for $\gamma_1=35.6^{\circ}$ the interferometer maps non-orthogonal states onto orthogonal ones at output 1. Further increment of $\gamma_1$ leads to an increment of the visibility until reaching the maximum value of one at $\gamma_1=45^\circ$, in which case the projection onto $|1\rangle$ is associated with a WPM with polarization $V$ that erases completely the WPI. The experimental results followed this behaviour and when the corrections for experimental imperfections were taken into account the agreement with theory became even better, as can be seen in the solid lines of figure~\ref{FIG4}(a). For $\gamma_1=35^{\circ}$ the visibility obtained was $0.17\pm 0.04$ that is close to the value expected for this  angle (0.14) which is not exactly the same one that would generate nearly orthogonal states. The worse case happened for $\gamma_1=45^{\circ}$ where the visibility obtained was just $0.73\pm 0.04$, when we expect to see at least a value close to 0.9.  This indicates  that there is still some WPI and we believe that the reason for that is the setting of the half-wave plate angle ($\gamma_1$). The slope of the visibility curve in figure~\ref{FIG4}(a) indicates that any small displacement from the right value of $\gamma_1=45^{\circ}$ implies a non-negligible variation of the visibility.

Figure~\ref{FIG1}(h) shows the erasure of the WPI for the experiment in \ref{FIG1}(e), where the WPM states were nearly orthogonal. We place a linear polarizer in front of the idler detector at output port 1. When the polarizer projects the idler polarization onto $V$ the interference is restored in the form of fringes. When the projection is onto $H$ the interference is restored in the form of antifringes. The averaged sum of these two patterns gives a pattern roughly equal to that of figure~\ref{FIG1}(e). 

\begin{figure}[t]
\hspace{2.5cm}
\rotatebox{-90}{\includegraphics[width=0.6\textwidth]{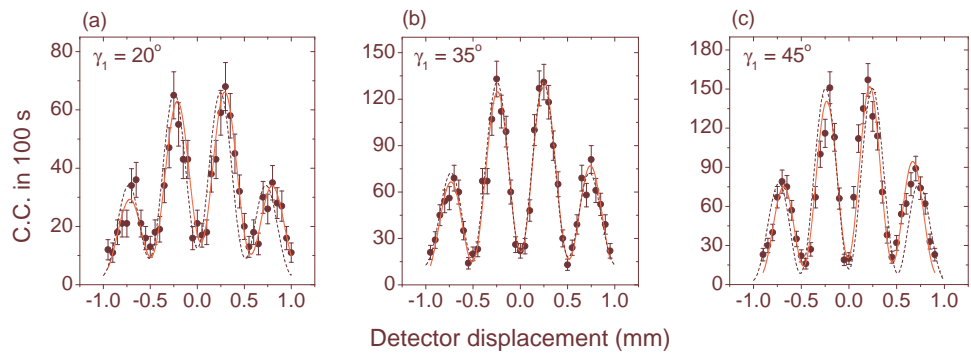} }
\vskip -4.3cm
 \caption{Interference patterns measuring the idler at output port 2 (figure~\ref{FIG0}) keeping  $\gamma_2=20^\circ$ and  (a)  $\gamma_1=20^\circ$, (b) $\gamma_1=35^\circ$, (c) $\gamma_1=45^\circ$. These graphics might be compared with those in figures~\ref{FIG1}(c), (e) and (g), respectively. Dashed lines correspond to the theoretical prediction based on equation~(\ref{PATTERNS}). Solid lines are the theoretical predictions with corrections described in the text.} 
 \label{exit2} 
\end{figure}

Interference patterns for measurements at output port 2 are shown in figure~\ref{exit2}. The antifringes are observed because the inner product in equation~(\ref{ALPHA2}) for the values of $\gamma_1$ shown in the insets are positive while $\phi=\pi$ in equation~(\ref{PATTERNS}) \cite{Comment2}. The patterns (a)--(c) might be compared with those of figures~\ref{FIG1}(c), (e) and (g), respectively. The obtained visibilities are: (a) $0.66\pm 0.4$, (b) $0.73\pm 0.3$ and (c) $0.77\pm 0.3$, which are in good agreement with equation~(\ref{ALPHA2}) when you consider the corrections described previously. 

As shown in equation~(\ref{cnorm}) the initial inner product of the WPM states ($\alpha$) might be preserved. Here this quantity is $0.7$ and the same value is obtained substituting the parameters $|a|$, $|b|$ and $\gamma_2$ in the right side of equation~(\ref{cnorm}) as a function of $\gamma_1$. Using the experimentally measured probabilities, visibilities as well as the phase of the patterns (which determine the sign of the inner product) we calculate the right side of equation~(\ref{cnorm}) for the three related measurements shown in figures~\ref{FIG1} and \ref{exit2}. The obtained values are shown in table~\ref{tab:cons}. Although far from the expected value, when we take into account the limits and errors of the experiment\footnote{In this case the right side of equation~(\ref{cnorm}) is calculated from the values (solid lines) used to fit the experimental results shown in  figures~\ref{FIG4}(a) and (b) for $|\alpha^{(1)}(\Psi_\mathrm{pol},\gamma_1,\gamma_2)|$ and $N_1$, respectively, and similarly for the quantities $|\alpha^{(2)}(\Psi_\mathrm{pol},\gamma_1,\gamma_2)|$ and $N_2$.}, we see that the experimental results are in good agreement with the values that could be reached, as can be seen in table~\ref{tab:cons}.

\begin{table}[t]
\caption{Conservation of inner product. The values shown here are calculated from the right side of equation~(\ref{cnorm}).}
\begin{indented}
\item[]
\begin{tabular}{@{}lllllll}
\br
HWP1  & $\;\;$  & Expected &  $\;\;$  &  Theory with &$\;\;$ & Experimental  \\[-0.5mm]
angle $(\gamma_1)$  & $\;\;$  & value &  $\;\;$  & corrections &$\;\;$ & results  \\
\mr
$20^{\circ}$   && $0.70$  && $0.64$  && $0.65 \pm 0.09$ \\
$35^{\circ}$   && $0.70$  && $0.60$  && $0.58 \pm 0.09$ \\
$45^{\circ}$   && $0.70$  && $0.51$  && $0.56 \pm 0.08$ \\
\br
\end{tabular}
 \label{tab:cons}
\end{indented}
\end{table}

\section{Conclusion}   \label{sec_conclusion}

The quantum eraser is a phenomenon closely related with several important subjects lying at the heart of quantum theory: interference, complementarity, entanglement, non-locality,  delayed choice and state discrimination as well. Here we have mainly exploited this later aspect by implementing an experiment where an interfering single photon is entangled with another one serving as which-path marker and whose states are non-orthogonal. Consequently, we neither have complete WPI nor a null visibility of the interference pattern. However, we have shown that it is possible to apply a probabilistic method for unambiguous modification of the inner product of such states to achieve any desired value of this quantity. Therefore we could, for instance, map the non-orthogonal states onto orthogonal ones and get complete WPI (destroying the interference) or onto collinear states to recover the interference with maximum visibility. This would not be possible by using just projection analysis.

To perform this experiment we used photon pairs generated by SPDC with arbitrary degree of polarization entanglement. Sending one of these photons through a double slit we observed single-photon interference (by a suitable focusing of the pump beam into the crystals) showing that such interference is independent of the polarization degree of freedom. Placing quarter-wave plates with their fast axes orthogonally oriented behind each slit we were able of coupling the spatial and polarization degrees of freedom introducing WPI for the interfering photon. After a suitable projection of its polarization we made this WPI to be carried just by the spatially separated photon which then goes through a polarization-sensitive Mach-Zehnder interferometer that implements the protocol described above. In this setup it is possible, for a suitable setting of the interferometer, to achieve an complete WPI by detecting the WPM at one output port of the interferometer or maximum visibility interference if the detection is performed at the other port. The only constraint is that the initial inner product of the WPM states might be preserved since the unitary operation applied on the WPM is local, what has been shown here.

\ack
We would like to thank C. H. Monken for lending us the sanded quarter-wave plates used in the double slit. This work was supported by Grants Milenio ICM P06-067-F and FONDECyT  N$^\circ$1061046 and N$^\circ$1080383.


\begin{thebibliography}{99}
\bibitem{Comment2} During the alignment of the birefringent double slit in this experiment we set the phase $\phi$ to be $\pi$ in equation~(\ref{spa}) (and every subsequent equation where it appears).
\end{thebibliography}
\end{document}